\documentclass[pra,aps,reprint,letterpaper,nofootinbib,superscriptaddress,floatfix]{revtex4-2}
\usepackage[utf8]{inputenc}
\usepackage{times}
\usepackage{graphicx}
\usepackage{epsfig}
\usepackage{amsfonts}
\usepackage{amsmath}
\usepackage{amssymb}
\usepackage{dsfont}
\usepackage{subcaption}

\usepackage{amsthm}% in order to use a black box at the end of proofs
\usepackage{color,xcolor}
\usepackage[normalem]{ulem}
\usepackage{multirow}
\usepackage[colorlinks=true,linkcolor=blue,citecolor=blue,urlcolor=blue,breaklinks]{hyperref}
\usepackage[breaklinks]{hyperref}
\usepackage{bbm}
\usepackage[T1]{fontenc}
\usepackage{makecell}
\usepackage[normalem]{ulem}
\usepackage{multirow}
\usepackage{subcaption}
\usepackage[font=small,labelfont=bf,
   justification=justified,
   format=plain]{caption}
\begin{document}
\title{Three-qubit W state tomography via full and marginal state reconstructions on{\texttt{ ibm\_osaka}}}
\author{H. Talath}
\email{talathumera45@gmail.com}
\affiliation{Department of Physics, Bangalore University, Jnanabharathi, Bengaluru -560056, India}

\author{B. P. Govindaraja}
\email{govindarajabp@gmail.com}
\affiliation{Department of Physics, Kuvempu University, Shankaraghatta-577451, India}

\author{B. G. Divyamani}
\email{divyamanibg@gmail.com}
\affiliation{Tunga Mahavidyalaya, Thirthahalli 577432, India}

\author{Akshata Shenoy H.}
\email{akshata.shenoy@ug.edu.pl}
\affiliation{International Centre for Theory of Quantum Technologies, University of Gda{\'n}sk, 80-308 Gda{\'n}sk, Poland.}

\author{A. R. Usha Devi}
\email{ushadevi@bub.ernet.in}
\affiliation{Department of Physics, Bangalore University, Jnanabharathi, Bengaluru -560056, India}

\author{Sudha}
\email{tthdrs@gmail.com}
\affiliation{Department of Physics, Kuvempu University, Shankaraghatta-577451, India}

\begin{abstract}
We present a three-qubit quantum state tomography scheme requiring a set of  17 measurement  settings, significantly reducing the experimental overhead compared to the conventional 63 Pauli measurement  settings. Using IBM's 127-qubit open-access quantum processor \texttt{ibm\textunderscore{osaka}}, we prepare the three-qubit W state 
 and employ our tomography scheme to reconstruct it. Additionally, we implement a two-qubit tomography protocol, involving  7 measurement  settings, on \texttt{ibm\textunderscore{osaka}} to reconstruct {\em two} of the two-qubit marginals of the W state. This serves as a {\em proof-of-principle} demonstration of the well-known theoretical result that any two of the two-qubit reduced density matrices can uniquely determine most of the whole three-qubit pure states. We show that the fidelity of the W-state reconstructed from its two-qubit subsystems is consistently larger than that obtained from the full three-qubit tomography, highlighting the practical advantage of the subsystem-based tomography approach.
\end{abstract}

\maketitle

\section{Introduction}
Quantum state tomography (QST) is a key technique for reconstructing quantum states and plays a vital role in benchmarking and validating the performance of quantum computing hardware~\cite{Cramer_2010}. The QST process involves a complete set of measurements on a number of identical copies of the quantum state to determine the real independent parameters of the state~\cite{Altepeter2004}. Noting that the number of independent parameters characterizing a $N$-qubit state is $(2^N)^2-1$, one needs as many Pauli measurements  for the corresponding state tomography. In other words an exponentially scaling resources are required for the realization of QST based on Pauli scheme. It is thus a challenge to find efficient tomographic schemes which require lesser number of measurements for reconstructing a $N$-qubit state. There have been continued efforts to improve the efficiency of quantum tomographic schemes~\cite{TothQST, Flammia_2012, Vanner2013, Baldwin2016, DDA15, Xin2017,Schmale2022,Hu2024,Abo2024}. Another important class of efficient tomographic protocols relies on mutually unbiased bases (MUBs). It is well known that a complete set of MUBs in a $d$-dimensional Hilbert space consists of $d+1$ orthonormal bases~\cite{nielsen2002quantum, MUB1, MUB2}. For three qubits ($d=8$), this implies that $d+1$=9 MUBs suffice to perform informationally complete projective measurements for full state tomography. However, the implementation of these ideal measurement sets on current NISQ hardware is often constrained by the native gate set, qubit connectivity, and noise. Most experimental tomography — including superconducting qubits, trapped ions, silicon spin qubits, and photonic platforms — still employs Pauli-based protocols, entangled basis rotations, or overcomplete POVMs rather than the minimal $d+1$ MUB protocols.

The QST scheme of reconstructing the multiparty state using its reduced density matrices has been explored extensively~\cite{DDA15,Xin2017}. This scheme evidently leads to reduction of quantum resources required for the task.  
It is pertinent to point out that the method of {\em determining the whole from its parts} is related to the quantum marginal problem, where the possibility of unique determination of a whole quantum state with the help of a set of reduced density matrices is investigated~\cite{C63}. Considerable research has been carried out to address the question "is it  possible to determine the higher order quantum correlations completely and uniquely from the lower order ones?"~\cite{LPW02,LW2002-PartsDetWhole,jones2005parts,WalckLyons2008,Parashar_2009,Devi_2011}. To this end, we focus on the specific result that {\em almost every pure three-qubit state, except the {\rm GHZ} state, can be determined completely by two of its two-qubit reduced density matrices}~\cite{LPW02,WalckLyons2008}. Diósi~\cite{diosi2004} has developed an explicit procedure to uniquely reconstruct a three-party pure state from {\emph {any two}} of its two-party reduced density matrices.  

We consider the three-qubit W state:
\begin{equation}
\vert {\rm W}_{ ABC}\rangle=\frac{\vert 1_A0_B0_C\rangle +\vert 0_A1_B0_C\rangle+\vert 
0_A0_B1_C\rangle}{\sqrt{3}}
\end{equation}
which forms an important class of permutation symmetric states~\cite{JQC07,AP06} exhibiting robustness against noise~\cite{C22} and loss of qubits~\cite{DVC00}. 
%It is shown~\cite{Parashar_2009} that they can be determined uniquely from two of its two-qubit reduced density matrices.
Here we employ two different QST schemes to reconstruct the three-qubit state $\vert \rm W_{ABC}\rangle$ 
experimentally in IBM's  open-access quantum processor \texttt{ibm\textunderscore{osaka}}:  (i) A three-qubit QST  scheme consisting of   17 measurement  settings proposed here to reconstruct any arbitray three-qubit state.    (ii) A two-qubit tomography protocol involving  7 measurement  settings  to reconstruct {\em two} of the two-qubit reduced states (marginals) of the three-qubit $\rm \vert W_{ABC}\rangle$ state. This QST scheme is employed as an alternative approach to determine the $\rm \vert W_{ABC} \rangle$ state on the \texttt{ibm\textunderscore{osaka}} processor.  Our work demonstrates an experimental implementation of the whole-from-parts protocol, i.e., reconstructing a global pure three-qubit quantum state from its two-qubit marginals on the IBM quantum platform.

%through its experimentally obtained subsystems $\rho_{AB}~=~{\rm Tr}_C\left(\vert {\rm W}_3\rangle\langle {\rm W}_3\vert\right)$, 
%$\rho_{BC}={\rm Tr}_A\left(\vert {\rm W}_3\rangle\langle {\rm W}_3\vert\right)$.  
%We have made use of a two-qubit tomography scheme~\cite{sdogr:meln:para} for the experimental determination of two-qubit subsystems$\rho_{AB}$, $\rho_{BC}$ of the W-state $\vert W_3\rangle$ on the . The whole pure state $\vert {\rm W}_3\rangle^{\rm parts}$ determined from its subsystems using the Diosi protocol~\cite{l:diosi} is found to have better fidelity with the W-state 
%$\vert {\rm W}_3\rangle$.  
  
We have organized the paper as follows: In Section~\ref{sec:whparts}, we provide an overview of key results on the determination of a global pure qubit state from its reduced density matrices. We also outline the  protocol developed by Di{\'o}si~\cite{diosi2004} for reconstructing  three-party pure states using {\em two} of its two-party reduced density matrices in subsection~\ref{subsec:Diosi}. 
In Section~\ref{sec:QST Scheme}, we present a QST scheme for arbitrary three-qubit states that requires only 17 measurement  settings. We also describe a two-qubit tomography protocol involving just 7 measurement  settings, which we employ to reconstruct two of the reduced two-qubit density matrices. 
Section~\ref{sec:experiment} discusses details of  experimental preparation of the state $\vert {\rm W}_{ABC} \rangle$, the implementation of both the two-qubit and three-qubit QST protocols on the IBM quantum processor \texttt{ibm\textunderscore{osaka}}, 
and the results obtained.  Finally, a summary of our results is given in Sec.~\ref{sec:Summay}.     

\section{Determining the Whole from Its Parts}
\label{sec:whparts}
Understanding the extent to which a multipartite quantum state is determined by its subsystems is a fundamental question with wide-reaching implications in quantum information, many-body physics, and the foundations of quantum mechanics. A central theme is whether the correlations present in a global state can be entirely inferred from those among fewer parties~\cite{C63, LW2002-PartsDetWhole, LPW02, linden2002high, jones2005parts,WalckLyons2008, Parashar_2009, Devi_2011, DDA15, Xin2017}. Linden and Wootters~\cite{LW2002-PartsDetWhole} showed that the reduced density matrices of about two-thirds of the parties are sufficient to uniquely determine most generic pure states of an $N$-qubit system. Jones and Linden~\cite{jones2005parts}  strengthened this result by proving that even slightly more than $N/2$ subsystems suffice to determine a generic $N$-party pure state. Parashar and Rana~\cite{Parashar_2009} established that $N$-qubit  W states, which represent a distinct class of entangled states with genuine multipartite entanglement, are also uniquely determined by their bipartite marginals. In an extended context, some of the present authors have shown that permutation invariant $N$-qubit states belonging to the Dicke class can be uniquely reconstructed from just two of their $N-1$-qubit marginals~\cite{Devi_2011}. Furthermore, Walck and Lyons~\cite{WalckLyons2008,Walck_2009} identified that the class of GHZ states and their local unitary equivalents  possess irreducible global correlations that cannot be inferred from any marginal systems. 

The theoretical result that almost all pure multiqubit states are uniquely determined by their lower-order marginals has been experimentally validated using nuclear magnetic resonance (NMR) quantum information processors: Dogra \textit{et al.}~\cite{DDA15} demonstrated the reconstruction of generic three-qubit pure states from two of their bipartite marginals, while Xin \textit{et al.}~\cite{Xin2017}  reconstructed three- and four-qubit pure states from reduced two- and three-qubit density matrices. Both experiments reinforce the foundational principle that, for generic pure states, the parts uniquely determine the whole.

Focusing on the specific case of three-qubit systems, it was first theoretically shown by Linden, Popescu, and Wootters~\cite{LPW02} that pure three-qubit states are uniquely characterized by their two-qubit reduced density matrices.  Diósi~\cite{diosi2004} developed a theoretical protocol for reconstructing a generic pure state $\rho_{ABC} = \vert \Psi_{ABC} \rangle \langle \Psi_{ABC} \vert$ of a {\em three-party} system from any two of its bipartite reduced density matrices — $\rho_{AB}$, $\rho_{BC}$, or $\rho_{AC}$.

In the following, we outline the Dio\'si protocol~\cite{diosi2004} illustrating how we could employ it to reconstruct the three-qubit $\vert {\rm W}_{\rm ABC}\rangle$ state  from {\em any two} of its two-qubit marginals. 

\subsection{Construction of three-party pure state from its subsystem density matrices}
\label{subsec:Diosi}
%It is known that all three-qubit states except the GHZ class can be reconstructed uniquely and fully from any two of its two-qubit density matrices \cite{LPW02}. Furthermore these can be obtained by performing two-qubit tomography using $7$ complete set of measurements as opposed to $15$ Pauli measurements \cite{DDA15}. Finally, using the protocol of \cite{diosi2004}, the entire three-qubit state can be reconstructed from its reduced density matrices. This forms the second part of our work.

In this subsection we describe the procedure~\cite{diosi2004} to construct a pure three-party state from {\em two} of its two-party reduced states.  
\begin{itemize}
\item Given two bipartite subsystems $\rho_{AB}$ and $\rho_{BC}$ of a global tripartite state $\rho_{ABC}$, the single-party reduced states can be readily obtained as follows:
\begin{eqnarray*}
\rho_A &=& \mathrm{Tr}_B\,\rho_{AB}, \ \rho_B = \mathrm{Tr}_A \rho_{AB}, \ \rho_C = \mathrm{Tr}_B\,\rho_{BC}.
\end{eqnarray*}
\item When the global state  is a pure state $|\Psi_{ABC}\rangle$, the reduced states $\rho_A$ and $\rho_C$ respectively share their non-zero eigenvalues with $\rho_{BC}$ and $\rho_{AB}$.
\item Let the eigenvalues and corresponding normalized eigenvectors of the single-party marginals be denoted as:
\begin{eqnarray*}
\rho_A: && \ (\lambda_A^i,\, |i;A\rangle), \ 
\rho_B: \ (\lambda_B^j,\, |j;B\rangle), \\ 
&& \ \ \ \ \ \ \rho_C: \ (\lambda_C^k,\, |k;C\rangle),
\end{eqnarray*}
and let $|i;BC\rangle$ and $|k;AB\rangle$ denote the normalized eigenvectors of $\rho_{BC}$ and $\rho_{AB}$, respectively.
\item Since the three-party  state $\rho_{ABC}=|\Psi_{ABC}\rangle\langle \Psi_{ABC}|$  is pure, we have the following spectral decompositions:
\begin{eqnarray*}
\rho_A &=& \sum_i \lambda_A^i\, |i;A\rangle \langle i;A|, \\ 
\rho_{BC} &=& \sum_i \lambda_A^i\, |i;BC\rangle \langle i;BC|
\end{eqnarray*}
and 
\begin{eqnarray*}
\rho_C &=& \sum_k \lambda_C^k\, |k;C\rangle \langle k;C|, \\ 
\rho_{AB} &=& \sum_k \lambda_C^k\, |k;AB\rangle \langle k;AB|. 
\end{eqnarray*}
\item The structure of the pure state $|\Psi_{ABC}\rangle$  compatible with the pair $(\rho_A, \rho_{BC})$ is given by:
$$
|\Psi_{ABC}; \alpha\rangle = \sum_i e^{i\alpha_i} \sqrt{\lambda_A^i}\, |i;A\rangle \otimes |i;BC\rangle.
$$
 On the other hand, $\rho_{ABC}$ compatible with the the subsystem pair $(\rho_C, \rho_{AB})$ can be expressed as 
$$|\Psi_{ABC}; \gamma\rangle = \sum_k e^{i\gamma_k} \sqrt{\lambda_C^k}\, |k;AB\rangle \otimes |k; C\rangle,$$
where $\alpha = \{\alpha_i\}$ and $\gamma = \{\gamma_k\}$ are undetermined phase factors.
\item Since the whole pure state $|\Psi_{ABC}\rangle$ must be consistent with both decompositions, there must exist at least one choice of $\{\alpha_i\}$ and $\{\gamma_k\}$ such that:
$$
|\Psi_{ABC}; \alpha\rangle = |\Psi_{ABC}; \gamma\rangle \equiv |\Psi_{ABC}\rangle.
$$
Therefore, determining either the set $\{\alpha_i\}$ or $\{\gamma_k\}$ that satisfies the above condition suffices to uniquely construct the global pure state $|\Psi_{ABC}\rangle$.
\item Diósi~\cite{diosi2004} expressed the phase factors $\alpha_i$ as 
\begin{eqnarray*}
\alpha_i &=& \sum_j \left(\mathcal{A}^i_{jk}\right)^* \mathcal{C}^k_{ij}, \\ 
\mathcal{A}^i_{jk} &=& \langle jk | i;BC \rangle, \ \ 
\mathcal{C}^k_{ij} = \langle ij | k;AB \rangle. 
\end{eqnarray*}
where $\vert ij\rangle\equiv \vert i;A\rangle \otimes \vert j; B\rangle$ and $\vert jk\rangle\equiv \vert j;B\rangle \otimes \vert k; C\rangle$
These phase factors $\alpha_i$ specify the global pure state $|\Psi_{ABC}\rangle$  uniquely. 
\end{itemize}
It may be noted  that the Diósi procedure applies to any general three-party pure states $|\Psi_{ABC}\rangle$, where $A$, $B$, and $C$ are finite-dimensional systems. In the special case of three-qubit pure states, the indices $i$, $j$, and $k$ take values $0, 1$.

%In Section~IV  we employ the Diósi method  eigenvalues and corresponding normalized eigenvectors of the experimentally obtained %two-qubit subsystems  $\rho_{AB}^{\rm expt}$, $\rho_{BC}^{\rm expt}$ and their single qubit marginals. Using these, we evaluated %the coefficients ${\cal A}_{jk}^i$, ${\cal }C_{ij}^k$, $i,\,j,\,k=1,\,2$ and the phase factors 
%$\alpha_i$, i=1,\,2~(see %%(\ref{phase1})). The `whole' pure state %$\vert \rm{W}_3\rangle^{\rm expt}$  is arrived at using (\ref{sd1}), in all the five trials.	
\section{QST schemes to reconstruct three-qubit state} 
\label{sec:QST Scheme} 
A general three-qubit density matrix $\rho_{ABC} ~=~ \displaystyle\sum_{i,j,k,l,m,n=0,1} \rho_{ijk;lmn})\vert i,j,k\rangle\langle l,m,n\vert$ is a  hermitian, positive semidefinite $2^3~\times~2^3$  matrix,
%\begin{widetext}
%\begin{widetext}
 %\begin{equation}
 %\rho_{ABC} =  \begin{pmatrix}
  %      \rho_{000;000} & \rho_{000;001} & \rho_{000;010} & \rho_{000;011} & \rho_{000;100} & \rho_{000;101} & \rho_{000;110} & \rho_{000;111} \\
%\rho^*_{000;001} & \rho_{001;001} & \rho_{001;010} & \rho_{001,011} & \rho_{001,100} & \rho_{001,101} & \rho_{001,110} & %\rho_{001,111} \\
 %        \rho^*_{000;010} & \rho^*_{001;010} & \rho_{010;010} & \rho_{010;011} & \rho_{010;100} & \rho_{010;101} & \rho_{010;110} & \rho_{010;111} \\
 %        \rho^*_{000;011} & \rho^*_{001;011} & \rho^*_{010;011} & \rho_{011;011} & \rho_{011;100} & \rho_{011;101} & \rho_{011;110} & \rho_{011;111} \\
  %       \rho^*_{000;100} & \rho^*_{001;100} & \rho^*_{010;100} & \rho^*_{011;100} &  \rho_{100;100} & \rho_{100;101} & \rho_{100;110} & \rho_{100;111} \\
   %      \rho^*_{000;101} & \rho^*_{001;101} & \rho^*_{010;101} & \rho^*_{011;101} & \rho^*_{100;101} & \rho_{101;101} & \rho_{101;110} & \rho_{101;111} \\
    %     \rho^*_{000;110} & \rho^*_{001;110} & \rho^*_{010;110} & \rho^*_{011;110} & \rho^*_{100;110} & \rho^*_{101;110} & \rho_{110;110} & \rho_{110;111} \\
     %    \rho^*_{000;111} & \rho^*_{001;111} & \rho^*_{010;111} & \rho^*_{011;111} & \rho^*_{100;111} & \rho^*_{101;111} & \rho^*_{110;111}& \rho_{111;111} \\
     %    \end{pmatrix}
%\end{equation} \label{3rho}
%\end{widetext}
with the real diagonal elements  obeying the unit trace condition $\sum_{i,j,k=0,1} \rho_{ijk;ijk}=1.$
The density matrix $\rho_{ABC}$ is characterized by a total of  $(2^3)^2 - 1 = 63$  real independent parameters in general. QST enables reconstruction of the state  from a complete set of 63 Pauli measurement  settings~\cite{james2001measurement}.  We present here 17 measurement  settings, suitable for implementation on a quantum processor, to determine a generic three-qubit quantum state. This QST scheme involves  single- and two-qubit unitary gates~\cite{nielsen2002quantum}, namely $\{ H, \text{CNOT}, R_{x}(\pi/2)\}$.   The complete set of measurement settings, along with the real and imaginary parts of the density matrix elements determined by each, is explicitly listed in Table~\ref{tab:3qtom}.

The measurement settings listed in Table~\ref{tab:3qtom} involve combinations of single - and two-qubit unitary gates — such as  $H$, $\text{CNOT}$, and $R_x(\pi/2)$ — applied  prior to projective measurement in the computational basis. Note that  measurements of Pauli $X,\ Y$ gates on a quantum processor requires one to make use of the relations $X~=~H\,Z\,H^\dag$, 
$Y~=~-R_x(\pi/2)\,Z\,R^\dag_x(\pi/2)$  so as to  map them onto measurable computational $Z$-basis. 
%\item 

Measurement of the three-qubit observables  $M_{ABC}~=~M_A\otimes M_B\otimes M_C$ registers 8 outcomes, labeled by $\{ijk, \ i,j,k=0,1\}$, and occurs with  probabilities 
\begin{eqnarray*}
{\rm P}_{M_AM_BM_C}(i,j,k)&=&\mbox{Tr}\,\left[\rho_{ABC}\left(\Pi_{M_A}(i)\otimes\Pi_{M_B}(j) \right.\right. \\ 
&&  \ \ \ \ \left.\left. \hskip 0.5in \otimes\Pi_{M_C}(k)\right)\right]
\end{eqnarray*}
where $\Pi_{M}(i)$ denote the eigen-projector of the  qubit observable $M$. We have denoted ${\rm P}_{CM_AM_BM_C}(i,j,k)$ as outcome probabilities wherever two-qubit  CNOT gates are involved in the measurement setting.      

%\item  
Note that  $I\otimes I\otimes I$ corresponds to measurement in the computational  ($Z$) basis for all the three qubits and it records  eight outcomes $\{ijk, \ i,j,k=0,1\}$ with probabilities ${\rm P}_{ZZZ}(i,j,k)$. This set of probabilities determine the diagonal elements 
$\{\rho_{ijk;ijk},i,j,k=0,1\}$  of the density matrix $\rho_{ABC}$.  The setting $H \otimes I \otimes I$ followed by $Z$ measurement on the qubits lead to probabilities ${\rm P}_{XZZ}(i,j,k)$ and they determine the real parts of selected off-diagonal terms (see Table~\ref{tab:3qtom}).  
 
 It may be pointed out  that ${\rm CNOT}_{\mu\nu}$ stands for the CNOT operation on the qubit pairs  $\mu\nu=AB, BC, AC$ and they indicate measurements  on the  state  ${\rm CNOT}_{\mu\nu}\,\rho_{ABC}\,{\rm CNOT}_{\mu\nu}$. For example, the measurement setting $(R_x(\pi/2) \otimes I \otimes I){\rm CNOT}_{AB}$   (see Table~\ref{tab:3qtom}) indicates rotation $R_x(\pi/2)$ on qubit $A$ of the three qubit state ${\rm CNOT}_{AB}\,\rho_{ABC}\,{\rm CNOT}_{AB}$ followed by computational basis measurements on all three qubits. The outcome probabilities are denoted by ${\rm P}_{CYZZ}(i,j,k)$. 
%\end{itemize}

While the three-qubit QST scheme based on 17 measurement  settings provides a direct  method for experimentally determining all the elements of a generic three-qubit state, our objective here is to explore an alternative approach: reconstructing the global three-qubit pure state from two of its two-qubit reduced density matrices. To this end, we employ a two-qubit QST protocol~~\cite{DMP21} consisting of  7 measurement  settings per marginal two-qubit subsystem.  We apply this QST scheme (see Table~\ref{tab:2qtomo}) to experimentally determine the reduced subsystem states $\rho^{\rm expt}_{AB}$ and $\rho^{\rm expt}_{BC}$ of the three-qubit W state. These marginals are then used, via the Diósi reconstruction protocol, to obtain the whole pure state $\vert W_{ABC}\rangle^{\rm expt}$ from its parts. The determination of the three-qubit W state from its two-qubit subsystems serves as a proof-of-principle demonstration of whole-state reconstruction from partial information. A detailed comparison is then made between the reconstructed W state (based on QST determining its two-qubit subsystems)  and the one obtained through full three-qubit QST to evaluate the consistency and effectiveness of the two approaches.

\section{Experimental Implementation on a Superconducting Quantum Processor}
\label{sec:experiment}
\begin{figure}[h]
    \centering
    \includegraphics[width=0.90\linewidth]{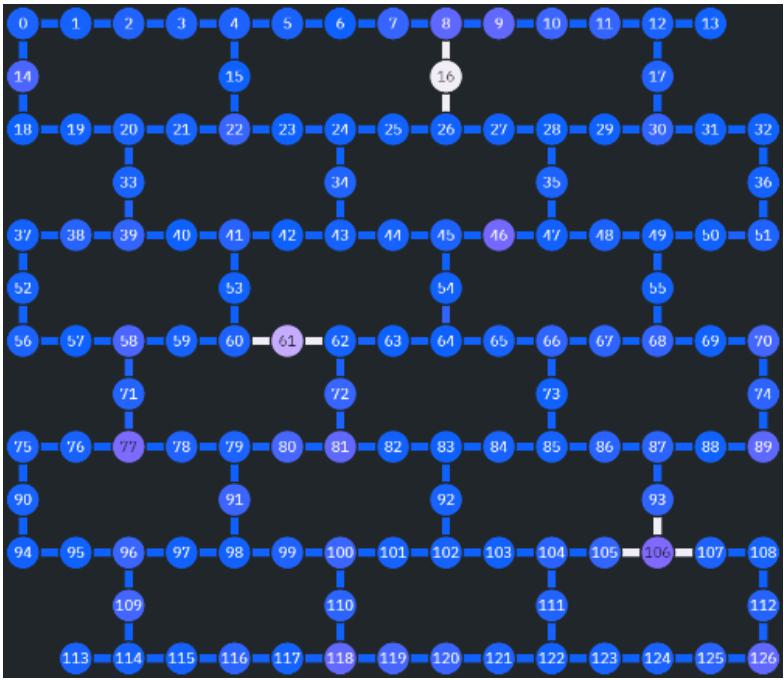}
    \caption{The architecture of 127-qubit IBM Quantum Processor \texttt{ibm\_osaka}. }
    \label{fig:layout}
\end{figure}
In this section, we describe the experimental implementation of our quantum state tomography protocols on the 127-qubit superconducting quantum processor \texttt{ibm\textunderscore{osaka}} accessed via the IBM Quantum cloud platform~\cite{ibmq}. The experiments were performed over a three-month period (May–July 2024), using qubits \texttt{q97}, \texttt{q98}, and \texttt{q99} of \texttt{ibm\textunderscore{osaka}}, which were selected for their consistently low readout and gate errors.  The physical layout of the device and the relevant qubit connectivity are illustrated in Figure~\ref{fig:layout}. Table~\ref{tab:calibration} lists the relevant calibration parameters (T1, T2, readout and gate errors) of the qubits \texttt{q97}, \texttt{q98}, and \texttt{q99} employed in the experiment. 

Our experimental data comprises five independent trials, each scheduled on different days and times to account for temporal variations in device performance. Each trial involved either 10,000 or 20,000 measurement shots per circuit, depending on backend availability and queue conditions. In total, the study involved approximately 70,000 measurement shots, providing a statistically robust dataset for reliable state reconstruction.

Each trial consisted of the following set of experiments:
\begin{itemize}
\item \textbf{17 quantum circuits} corresponding to the full three-qubit tomography protocol (see Table~\ref{tab:3qtom}),
\item \textbf{14 quantum circuits} (7 for each marginal) for two-qubit tomography of the reduced states $\rho_{AB}$ and $\rho_{BC}$ (see Table~\ref{tab:2qtomo}),
\item \textbf{6 calibration circuits} for readout error mitigation on qubits q97, q98, and q99.
\end{itemize}
Thus, each complete experimental trial involved a total of \textbf{31 tomography circuits and 6 calibration runs}. All experiments were executed using Qiskit's standard runtime environment, with appropriate transpilation levels applied to reduce circuit depth and mitigate gate errors.
\begin{widetext}
\begin{table}[ht]   
\centering
\captionsetup{width=0.8\textwidth}
\caption{A set of 17 settings used in the proposed three-qubit QST scheme and the elements $\rho_{ABC}$ determined using the probabilities of measurement outcomes}  
    \begin{tabular}{|c|c|c|}
        \hline
 & &  \\
     Settings & Probabilities $P_M(i,j,k)$  &    Elements of $\rho_{ABC}$ determined \\ 
     &&\\
     \hline
     $I \otimes I \otimes I$ & $P_{ZZZ}(i,j,k)$ &   $\{\rho_{ijk;ijk},i,j,k=0,1\}$ \\
     &&\\
     \hline
     $H\otimes I \otimes I$ & $P_{XZZ}(i,j,k)$ & Re$\rho_{000;100}$, Re$\rho_{001;101}$, Re$\rho_{010;110}$, Re$\rho_{011;111}$ \\
     \hline
       $I \otimes H \otimes I$ & $P_{ZXZ}(i,j,k)$ & Re$\rho_{000;010}$, Re$\rho_{001;011}$, 
       Re$\rho_{100;110}$, Re$\rho_{101;111}$ \\
    \hline
      $I \otimes I \otimes H$ & $P_{ZZX}(i,j,k)$ & Re$\rho_{000;001}$, Re$\rho_{010;011}$, 
       Re$\rho_{100;101}$, Re$\rho_{110;111}$ \\
      \hline
       $R_x\left(\frac{\pi}{2}\right) \otimes I \otimes I$ & $P_{YZZ}(i,j,k)$ & Im$\rho_{000;100}$, Im$\rho_{001;101}$, 
        Im$\rho_{010;110}$, Im$\rho_{011;111}$ \\
     \hline
     $I \otimes R_x\left(\frac{\pi}{2}\right) \otimes I$ & $P_{ZYZ}(i,j,k)$ & Im$\rho_{000;010}$, Im$\rho_{001;011}$, 
     Im$\rho_{100;110}$, Im$\rho_{101;111}$ \\
     \hline
     $I \otimes I \otimes R_x\left(\frac{\pi}{2}\right)$ & $P_{ZZY}(i,j,k)$ & Im$\rho_{000;001}$, Im$\rho_{010;011}$, 
      Im$\rho_{100;101}$, Im$\rho_{110;111}$ \\
     \hline
     $(H \otimes I \otimes I){\rm CNOT}_{AB}$ & $P_{CXZZ}(i,j,k)$ & Re$\rho_{000;110}$, Re$\rho_{001;111}$, 
      Re$\rho_{010;100}$, Re$\rho_{011;101}$ \\
     \hline
     $(I \otimes H \otimes I){\rm CNOT}_{BC}$ & $P_{CZXZ}(i,j,k)$ & Re$\rho_{000;011}$, Re$\rho_{001;010}$, 
      Re$\rho_{100;111}$, Re$\rho_{101;110}$ \\
     \hline
      $(H \otimes I \otimes I){\rm CNOT}_{AC}$ & $P_{CZZX}(i,j,k)$ & Re$\rho_{000;101}$, Re$\rho_{001;100}$,  
       Re$\rho_{010;111}$, Re$\rho_{011;110}$ \\
      \hline
      $(R_x\left(\frac{\pi}{2}\right) \otimes I \otimes I){\rm CNOT}_{AB}$ & $P_{CYZZ}(i,j,k)\}$ & Im$\rho_{000;110}$, Im$\rho_{001;111}$, Im$\rho_{010;100}$, Im$\rho_{011;101}$ \\
     \hline
     $(I \otimes R_x\left(\frac{\pi}{2}\right) \otimes I){\rm CNOT}_{BC}$ & $P_{CZYZ}(i,j,k)$ & Im$\rho_{000;011}$, Im$\rho_{001;010}$,  Im$\rho_{100;111}$, Im$\rho_{101;110}$ \\
     \hline
     $( R_x\left(\frac{\pi}{2}\right) \otimes I \otimes I){\rm CNOT}_{AC}$ & $P_{CZZY}(i,j,k)$ & Im$\rho_{000;101}$, Im$\rho_{001;100}$,  Im$\rho_{010;111}$, Im$\rho_{011;110}$ \\
     \hline 
          $(H \otimes H \otimes I){\rm CNOT}_{BC}$ & $P_{CXXZ} (i,j,k)$ & Re$\rho_{000;111}$, Re$\rho_{011;100}$ \\
   &&\\
     \hline
     $(R_x\left(\frac{\pi}{2}\right) \otimes R_x\left(\frac{\pi}{2}\right) \otimes I){\rm CNOT}_{BC}$ & $P_{CYYZ}(i,j,k)$ & Re$\rho_{001;110}$, Re$\rho_{010;101}$ \\
    &&\\
     \hline
      $(H \otimes R_x\left(\frac{\pi}{2}\right) \otimes I){\rm CNOT}_{BC}$ & $P_{CXYZ}(i,j,k)$ & Im$\rho_{011;100}$, Im$\rho_{000;111}$ \\
      &&\\
      \hline
      $(R_x\left(\frac{\pi}{2}\right) \otimes H \otimes I){\rm CNOT}_{BC}$ & $P_{CYXZ}(i,j,k)$ & Im $\rho_{001;110}$, Im $\rho_{010;101}$ \\
      \hline    
    \end{tabular}  \label{tab:3qtom} 
  \end{table}
\end{widetext}

%\begin{widetext}
\begin{table}[h]
 %\captionsetup{width=0.7\textwidth}
    \caption{Measurement settings used in the two-qubit tomography protocol}
\centering
    \begin{tabular}{|c|c|c|}
        \hline
Settings &   Probabilities   &  Two-qubit \\ 
&  $P_M(i,j)$ & density matrix elements  \\
    \hline
    $I \otimes I$ & $P_{ZZ} (i,j)$ & $\{\rho_{ij;ij}, i,j=0,1\}$ \\
    &&\\
    \hline
    $H \otimes I$ & $\{P_{XZ} (i,j)\}$ & Re$\rho_{00;10}$, Re$\rho_{01;11}$ \\
    &&\\
    \hline
    $I \otimes R_{x} (\pi/2)$ & $P_{ZY}(i,j)$ & Re$\rho_{00;01}$, Re$\rho_{10;11}$ \\
    &&\\
    \hline
    $I \otimes H$ & $P_{ZX} (i,j)$ & Im$\rho_{00;10}$, Im$\rho_{01;11}$ \\
    &&\\
    \hline
    $R_{x} (\pi/2) \otimes I$ & $\{P_{YZ} (i,j)\}$ & Im$\rho_{00;01}$, Im$\rho_{10;11}$ \\
    &&\\
    \hline    
    $(H \otimes I){\rm CNOT}$ & $P_{CXZ} (i,j)$ & Re$\rho_{00;11}$, Re$\rho_{01;10}$ \\
    &&\\
    \hline
    $(R_{x} (\pi/2) \otimes I){\rm CNOT}$ & $P_{YZ} (i,j)$ & Im$\rho_{00;11}$, Im$\rho_{01;10}$ \\
    &&\\
    \hline
    \end{tabular}
       \label{tab:2qtomo}
\end{table}
%\end{widetext}

\begin{figure}[h]
    \includegraphics[width=1\linewidth]{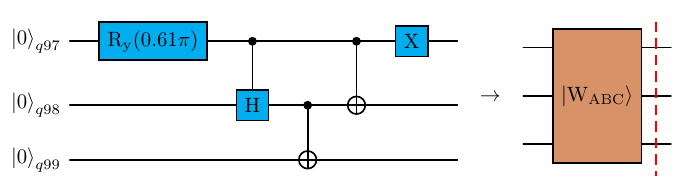}
    \caption{Quantum circuit used for the preparation of the three-qubit W state on \texttt{ibm\textunderscore{osaka}}.}
    \label{fig:circuit}
\end{figure}
\begin{figure} 
   	\includegraphics[width=4.0cm,height=2.5cm]{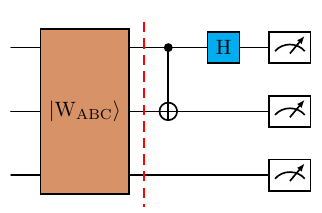}
	\includegraphics[width=4.5cm,height=2.5cm]{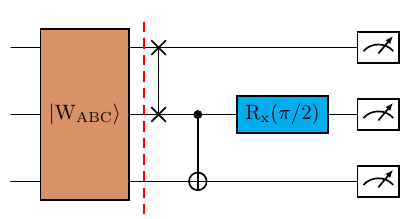}
    \hskip .75cm 
	\includegraphics[width=4.6cm,height=2.5cm]{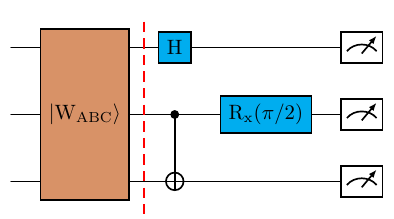}
     \caption{Circuits for the measurement settings  $(H\otimes I\otimes I){\rm CNOT}_{AB}, (R_x(\pi/2)\otimes I\otimes I){\rm CNOT}_{AC}$  and 
		$(H\otimes (R_x(\pi/2) \otimes I){\rm CNOT}_{BC}$.}
        \label{fig:qcir}
   	 \end{figure}
All three qubits were initialized in the computational basis state $\vert 0\rangle^{\otimes 3}$ , 
and the circuit shown in Fig.~\ref{fig:circuit} was applied to prepare the three-qubit W state $\vert {\rm W}_{ABC} \rangle$. 
Quantum circuits corresponding to selected measurement settings from the 17-measurement tomography scheme (see Table~\ref{tab:3qtom}) 
are illustrated in Fig.~\ref{fig:qcir}. In order to reconstruct the reduced density matrices $\rho_{AB}$ and $\rho_{BC}$, 
we performed measurements on qubit pairs \texttt{AB} = (\texttt{q97}, \texttt{q98}) and \texttt{BC} = (\texttt{q98}, \texttt{q99}), 
respectively (see Table~\ref{tab:2qtomo} for details of the two-qubit QST scheme). 
\begin{table}[h]
\caption{Calibration data for qubits \texttt{q97}, \texttt{q98}, \texttt{q99}, 
%and \texttt{q100} 
on the \texttt{ibm\textunderscore{osaka}} processor.} 
\begin{center}
    \begin{tabular}{|c|c|c|c|c|c|c|}
    \hline
      \thead{\textbf{Qubit}} & \textbf{\thead{T1 \\($\mu$s)}} & \textbf{\thead{T2 \\ ($\mu$s)}} & \textbf{\thead{ p(0|1) \\ error}} & \textbf{\thead{p(1|0) \\ error}} & \textbf{\thead{Readout \\ error}} & \textbf{\thead{CNOT \\error}}  \\ \hline
       97 & 493.76 & 408.28 & 0.023 & 0.008 & 0.015 & 97\textunderscore{98}:0.007 \\ \hline
       98 & 353.00 & 15.42 & 0.004 & 0.009 & 0.007 & 98\textunderscore91:0.004 \\ \hline
       99 & 313.09 & 204.00 & 0.030 & 0.034 & 0.032 & 99\textunderscore98:0.003 \\ \hline
       %100 & 231.04 & 220.50 & 0.012 & 0.019 & 0.015 & 100\textunderscore99:0.004 \\ \hline
    \end{tabular}
\end{center}
\label{tab:calibration}
\end{table}

\subsection{Error Mitigation}
\label{sec:mitigation}

IBM quantum processors, being part of the Noisy Intermediate-Scale Quantum (NISQ) era~\cite{Preskill_2018}, are susceptible to various errors arising from hardware imperfections and environmental interactions. Among the various noise sources, \emph{SPAM errors} — those arising from imperfect state preparation and measurement—are particularly significant. Measurement errors (also referred to as \emph{readout errors}) occur when the measurement outcome does not accurately reflect the eigenstate of the measured observable, while state preparation errors result in a deviation of the initialized state from the intended computational basis state. In IBM quantum processors, where $\rm Z$-basis measurements are default, there is a non-zero probability of detecting a $\vert 1\rangle$ outcome when the qubit is prepared in the $\vert 0\rangle$ state, and vice versa. These errors were characterized for each of the qubits \texttt{q97}, \texttt{q98}, \texttt{q99} of the processor and mitigated through a calibration-based strategy, as described below.\\
Let $p(i\vert j)$ denote the conditional probability of obtaining outcome $i$ when the qubit is prepared in state $\vert j\rangle$, for $i, j =0,1$. The measurement error on a single qubit can be modeled using the $2 \times 2$ calibration matrix $F$:
\begin{equation}
\label{eq:single_qubit_error}
F = \begin{pmatrix}
p(0\vert 0) & p(0\vert 1) \\
p(1\vert 0) & p(1\vert 1)
\end{pmatrix}
\end{equation}

Given an experimentally observed probability vector ${\rm P}_{M}^{\rm expt}$ (a column vector) for a single-qubit measurement $M$, the corresponding error-mitigated probability vector is computed as:
\begin{equation}
{\rm P}_{M}^{\rm mitigated} = F^{-1} {\rm P}_{M}^{\rm expt}
\end{equation}

For two- and three-qubit measurements, the error-mitigated probabilities are obtained using tensor products of the inverses of the respective calibration matrices:
\begin{eqnarray*}
{\rm P}_{M_AM_B}^{\rm mitigated} &=& \left( F_A \otimes F_B \right)^{-1} {\rm P}_{M_AM_B}^{\rm expt} \\
{\rm P}_{M_AM_BM_C}^{\rm mitigated} &= &\left( F_A \otimes F_B \otimes F_C \right)^{-1} {\rm P}_{M_AM_BM_C}^{\rm expt}
\end{eqnarray*}
These corrected probability distributions are subsequently used to estimate the elements of the density matrices in both the three-qubit and two-qubit tomography protocols.
\begin{figure}[h]
\begin{subfigure}[t]{0.5\textwidth}
    \includegraphics[width=1.0\linewidth]{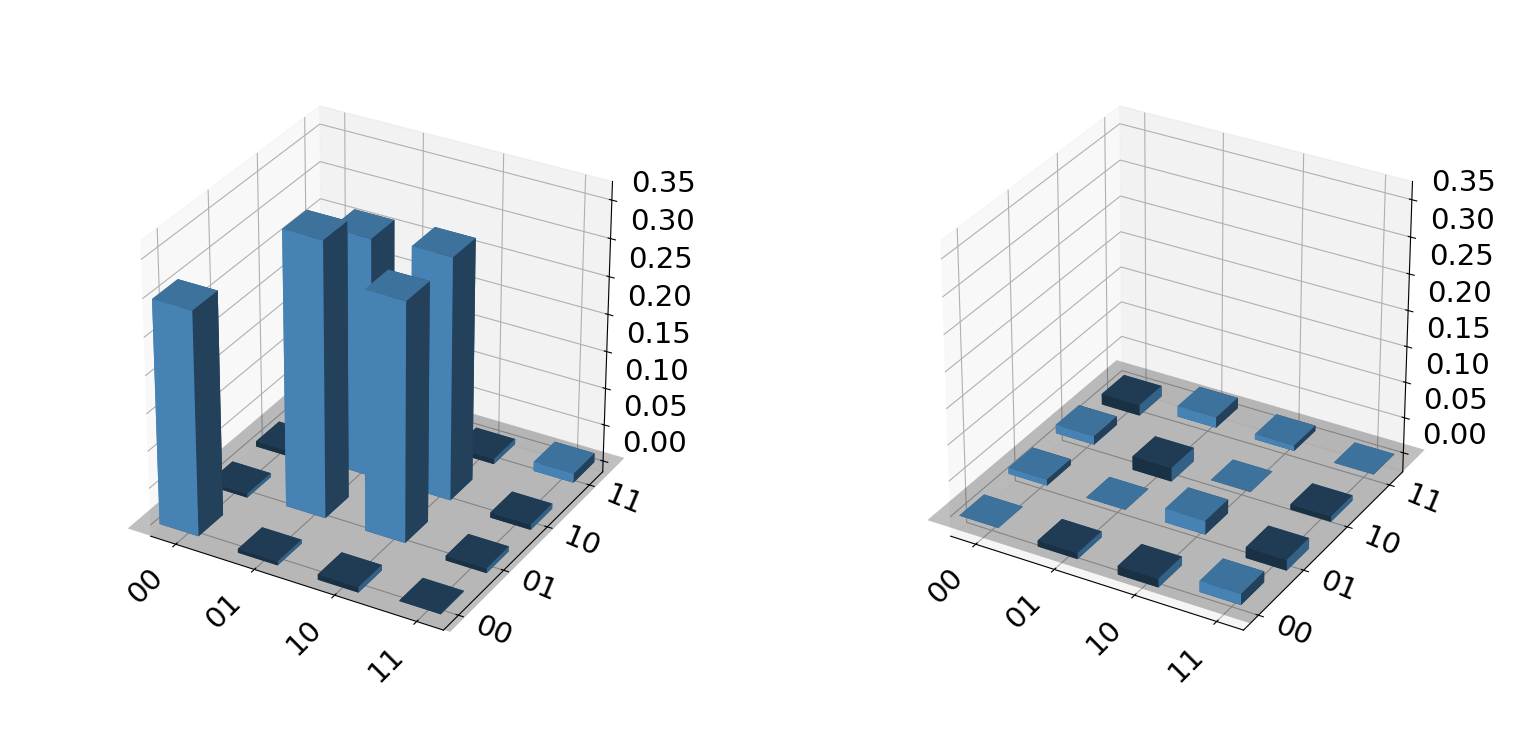}
    \caption{}
    \label{fig:a1}
\end{subfigure}
\begin{subfigure}[t]{0.5\textwidth}
    \includegraphics[width=1.0\linewidth]{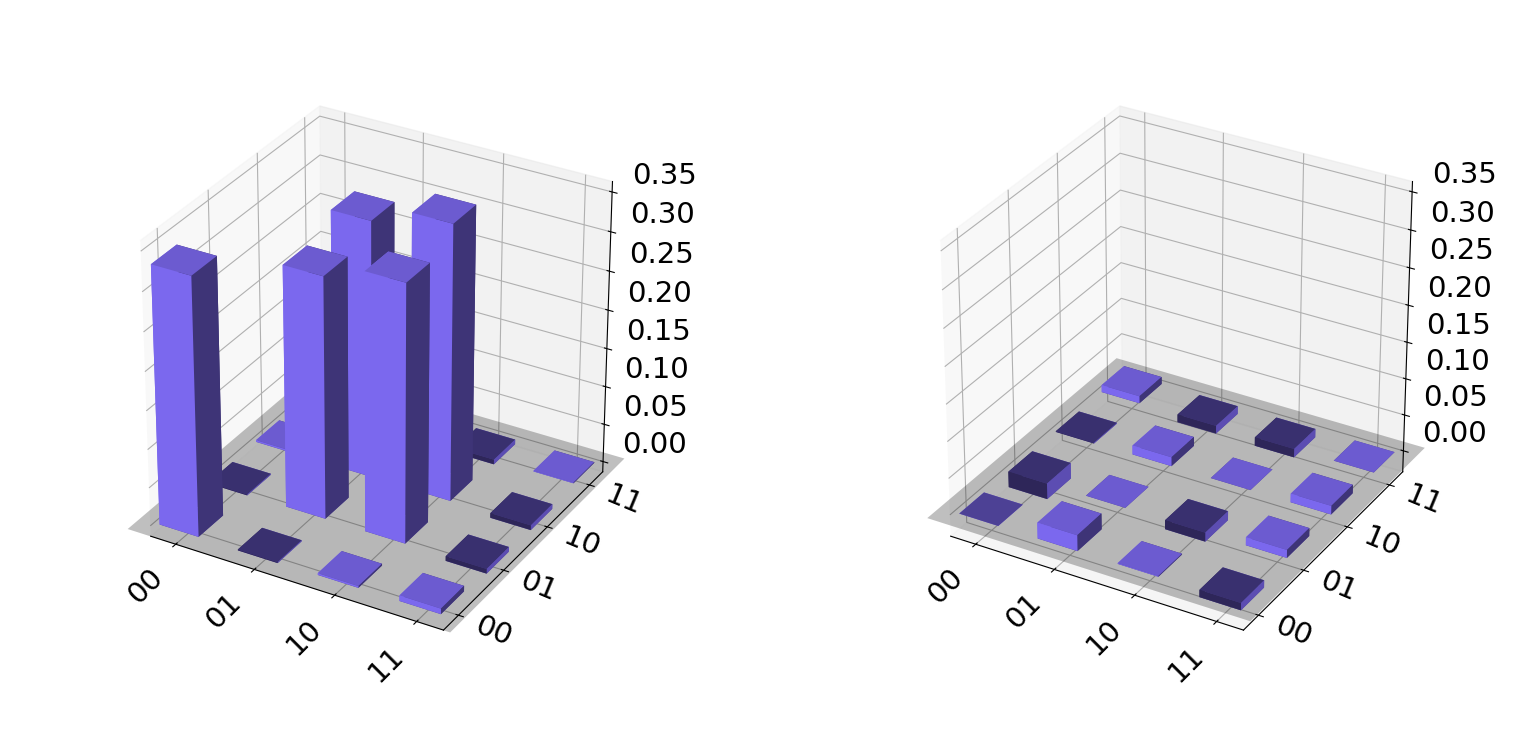}
    \caption{}
    \label{fig:a2}
\end{subfigure}
 \caption{Real (left) and imaginary (right) parts of the elements of the experimentally reconstructed two-qubit density matrices (a) $\rho^{\rm expt}_{AB}$, (b) $\rho^{\rm expt}_{BC}$ obtained from the two-qubit tomography scheme.}
    \label{fig:2qubit}
\end{figure}
While \emph{SPAM} error mitigation significantly improves the accuracy of quantum state tomography, additional errors intrinsic to NISQ hardware — such as gate infidelities and decoherence — can still degrade the quality of reconstructed states. Notably, even after correcting for readout errors, the resulting density matrices may exhibit small negative eigenvalues, thereby violating the physical requirement of positive semidefiniteness.

To remedy this, we adopt a spectral correction procedure as outlined in Ref.~\cite{Kaznady_2009}, which ensures the physical validity of the reconstructed density matrices. This involves (i) setting  negative eigenvalues of the reconstructed density matrix to zero (ii) renormalizing the  positive eigenvalues to satisfy the unit trace condition and reconstructing the physical density matrix via spectral decomposition using the renormalized positive eigenvalues and their corresponding eigenvectors. The final estimated state is then a valid physical density matrix obeying hermiticity, unit trace condition, and positive semidefiniteness.
\subsection{Reconstructed density matrices}

\begin{figure}
\begin{subfigure}[t]{0.5\textwidth}
    \includegraphics[width=1.0\linewidth]{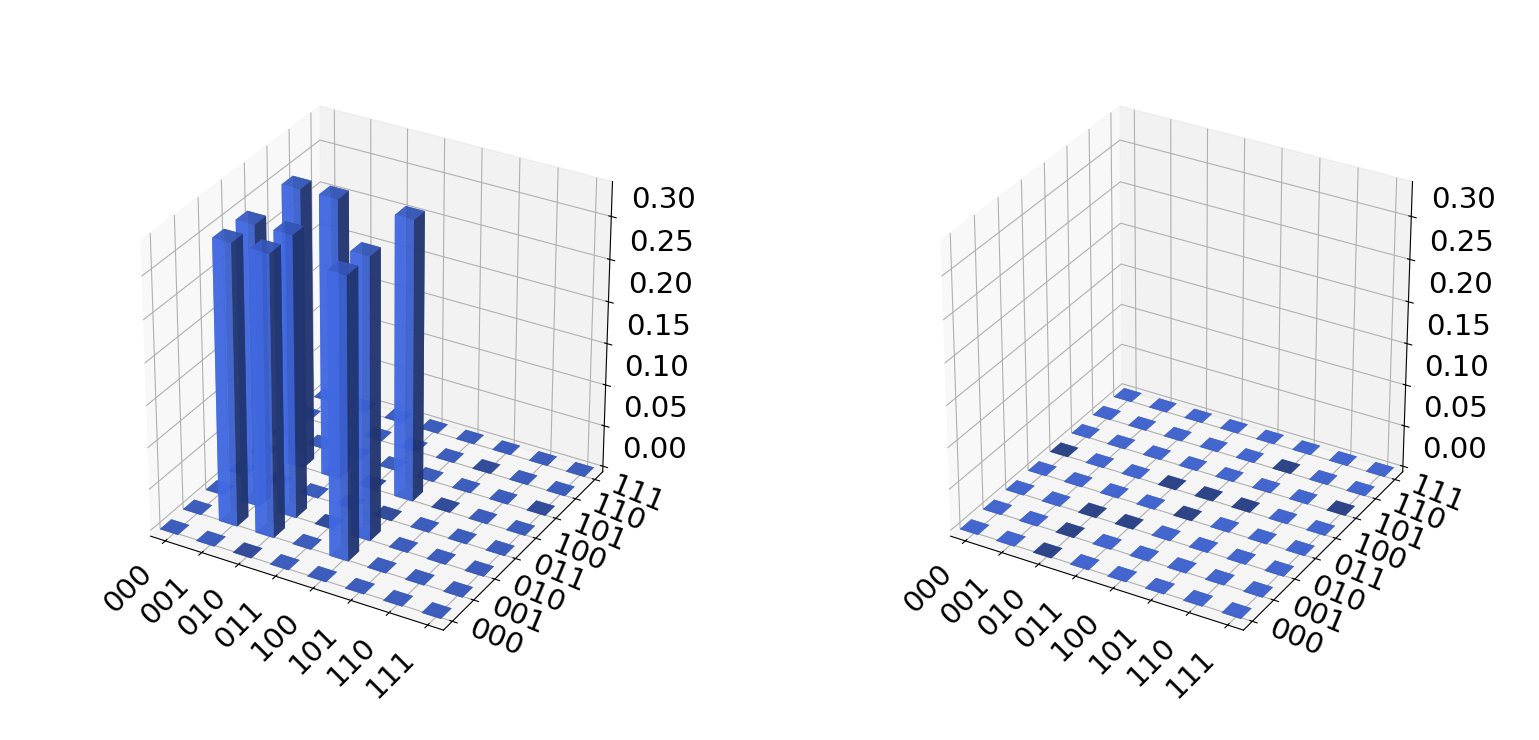}
    \caption{}
    \label{fig:a}
\end{subfigure}
\begin{subfigure}[t]{0.5\textwidth}
    \includegraphics[width=1.0\linewidth]{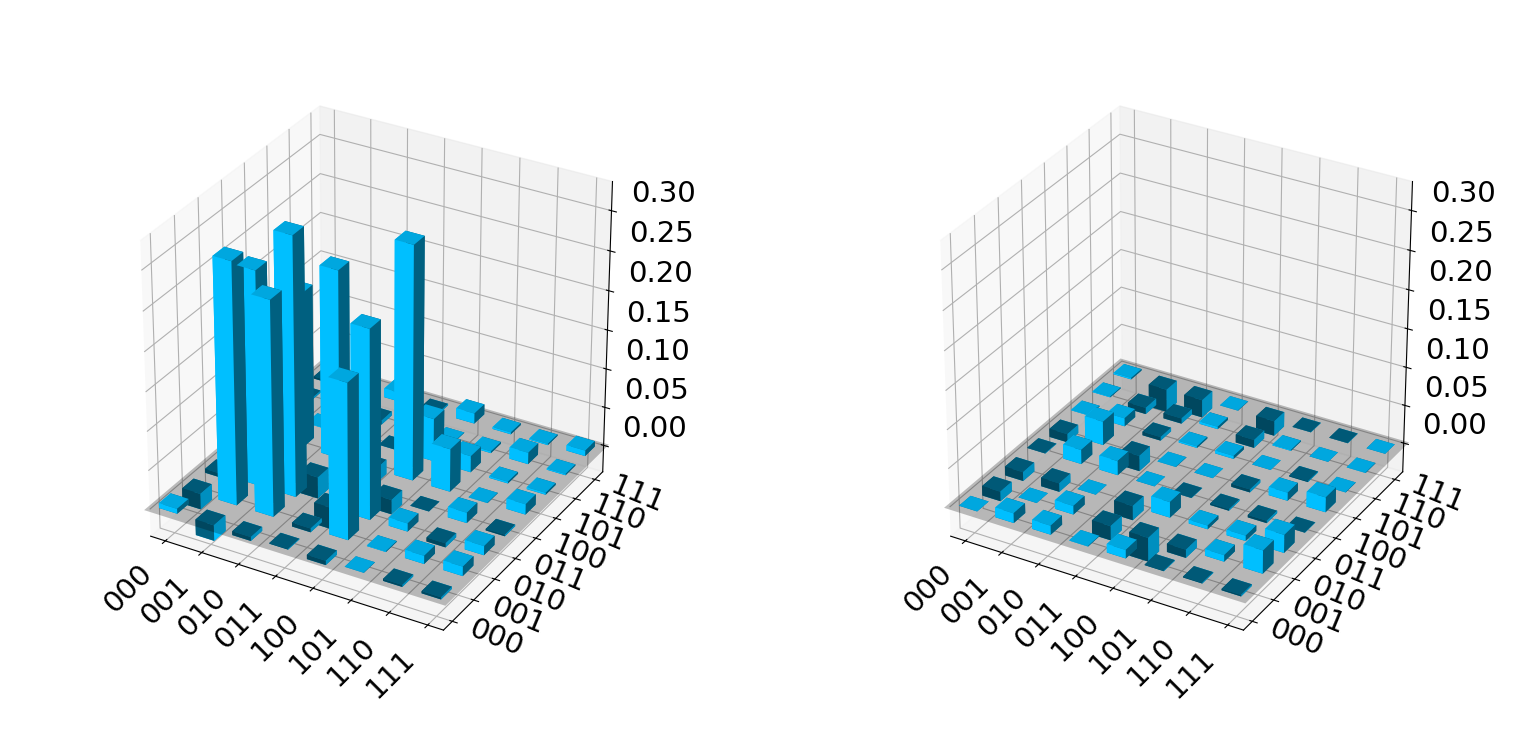}
    \caption{}
    \label{fig:b}
\end{subfigure}
\begin{subfigure}[t]{0.5\textwidth}
    \includegraphics[width=1.0\linewidth]{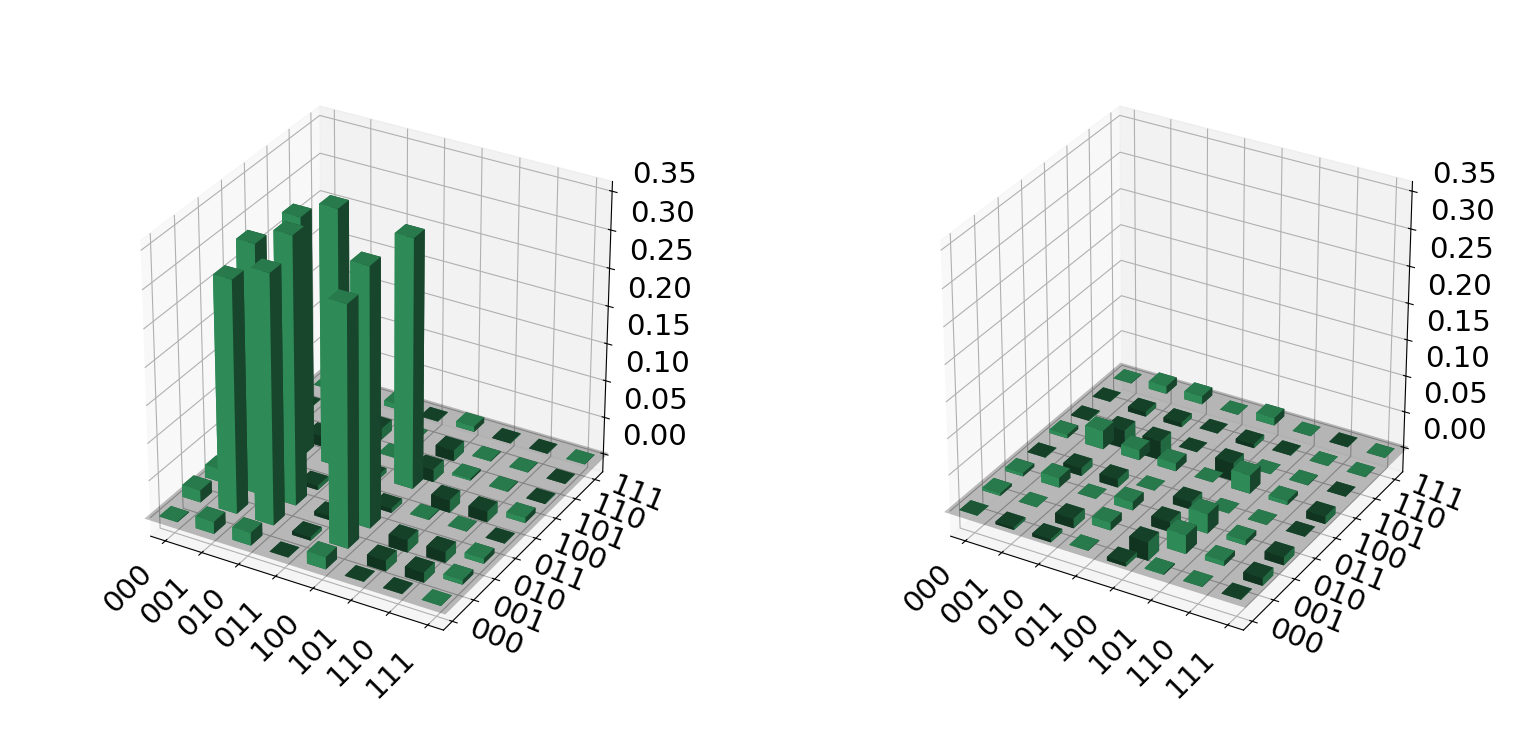}
    \caption{}
    \label{fig:c}
\end{subfigure}
    \caption{Real (right) and imaginary (left) parts of the elements of (a) the theoretical  three-qubit W state density matrix, (b) the experimentally reconstructed  $\rho_{ABC}^{I}$ obtained via full three-qubit tomography  (see Eq.~(\ref{rhoabcI})), (c)  $\rho_{ABC}^{II}$  (see Eq.~\ref{rhoabcII})) reconstructed  from the experimentally tomographed two-qubit marginals $\rho^{\rm expt}_{AB}$ and $\rho^{\rm expt}_{BC}$.} 
    \label{fig:thexpt}
\end{figure}

Using the experimental data and the tomography protocols described in Section~\ref{sec:QST Scheme}, we reconstruct (i) the whole three-qubit density matrix $\rho^{I}_{ABC}$ using the QST scheme with 17 measurement  settings. (ii) the reduced two-qubit states $\rho^{\rm expt}_{AB}$ and $\rho^{\rm expt}_{BC}$ using the tomography scheme containing  7-measurement  settings   (Table~\ref{tab:2qtomo}) \& the whole three-qubit state $\rho_{ABC}^{II}=\vert  W_{ABC}\rangle^{\rm expt}\langle W_{ABC}\vert$ reconstructed from them.

The following matrices correspond to one of the representative experimental tirals, after SPAM error mitigation and positivity correction (see Section~\ref{sec:mitigation}).
\begin{widetext}
\paragraph{ Two-qubit reduced density matrices:}
\begin{equation} \label{rhoabbc}
\rho^{\rm expt}_{AB} =
\begin{pmatrix}
0.31 & 0 & 0 & -0.02i \\
0 & 0.36 & 0.32 + 0.03i & 0 \\
0 & 0.32 - 0.03i & 0.31 & 0 \\
0.02i & 0 & 0 & 0
\end{pmatrix},
\qquad
\rho^{\rm expt}_{BC} =
\begin{pmatrix}
0.31 & -0.01i & 0 & 0 \\
0.01i & 0.31 & 0.33 + 0.01i & 0 \\
0 & 0.33 - 0.01i & 0.36 & -0.01i \\
0 & 0 & 0.01i & 0
\end{pmatrix}
\end{equation}
\paragraph{Three-qubit density matrix from full-state tomography:}
	{\footnotesize \begin{equation}
\label{rhoabcI}
		\rho_{ABC}^{I} =
		\begin{pmatrix}
			0 & -0.02 + 0.01i & 0.01i & 0 & 0.015i & 0 & 0 & 0 \\
			-0.02 - 0.01i & 0.31 & 0.27 + 0.01i & -0.02i & 0.20 - 0.02i & 0.05i & 0.01 + 0.01i & 0.01 + 0.03i \\
			-0.01i & 0.27 - 0.01i & 0.33 & -0.02 - 0.01i & 0.25 + 0.02i & 0.01 + 0.01i & 0.01i & 0.01 + 0.03i \\
			-0.01i & 0.02i & -0.02 + 0.01i & 0.01 & -0.02 & 0 & 0.01 & 0 \\
			-0.01i & 0.20 + 0.02i & 0.25 - 0.02i & -0.02 & 0.29 & -0.02 & 0.01i & 0.01 + 0.02i \\
			0 & -0.05i & 0.01 - 0.01i & 0 & -0.02 & 0.02 & 0 & 0 \\
			0 & 0.01 - 0.01i & -0.01i & 0.01 & -0.01i & 0 & 0.01 & 0 \\
			0 & 0.01 - 0.03i & 0.01 - 0.03i & 0 & 0.01 - 0.02i & 0 & 0 & 0
		\end{pmatrix}
	\end{equation} }
and
\paragraph{Whole three-qubit state reconstructed from two-qubit reduced marginals:}
\begin{equation}
\label{rhoabcII}
\rho_{ABC}^{II} =
\begin{pmatrix}
0 & 0 & 0 & 0 & 0 & 0 & 0 & 0 \\
0 & 0.31 & 0.34 - 0.01i & 0 & 0.31 - 0.04i & 0 & 0 & -0.01i \\
0 & 0.34 + 0.01i & 0.36 & 0.01i & 0.34 - 0.03i & 0.01i & 0 & -0.01i \\
0 & 0 & -0.01i & 0 & 0 & 0 & 0 & 0 \\
0 & 0.31 + 0.04i & 0.34 + 0.03i & 0 & 0.32 & 0 & 0 & -0.01i \\
0 & 0 & -0.01i & 0 & 0 & 0 & 0 & 0 \\
0 & 0 & 0 & 0 & 0 & 0 & 0 & 0 \\
0 & 0.01i & 0.01i & 0 & 0.01 + 0.01i & 0 & 0 & 0
\end{pmatrix}
\end{equation}
\end{widetext} 
 
The real and imaginary parts of the elements of the experimentally tomographed two-qubit density matrices $\rho^{\rm expt}_{AB}$ and $\rho^{\rm expt}_{BC}$ (see Eqs~(\ref{rhoabbc})) are displayed in Fig.~\ref{fig:2qubit}.

In Fig.~\ref{fig:thexpt}, the real and imaginary parts of the elements of  the theoretical density matrix $\rm \rho_{ABC}=\vert  W_{ABC}\rangle\langle W_{ABC}\vert$,   the density matrix $\rm \rho_{ABC}^{I}$ reconstructed from  the three-qubit QST scheme and the density matrix  $\rm \rho_{ABC}^{II}$ reconstructed from the experimentally tomographed two-qubit states are shown. 

We have evaluated the fidelities~\cite{nielsen2002quantum} 
\begin{eqnarray*}
F\left(\rho_{ABC},\rho^{I}_{ABC}\right)&=&\left(\langle W_{ABC} \vert \rho^{I}_{ABC}\vert W_{ABC}\rangle \right)^{1/2}   \\
F\left(\rho_{ABC},\rho^{II}_{ABC}\right)&=&\left(\langle W_{ABC} \vert \rho^{II}_{ABC}\vert W_{ABC}\rangle \right)^{1/2}
\end{eqnarray*}
between the theoretical three-qubit density matrix $\rho_{ABC}~\equiv~\vert W_{ABC}\rangle\langle W_{ABC}\vert$ and the corresponding experimentally reconstructed states $\rho^{I}_{ABC}$, $\rho^{II}_{ABC}$ for all five  experimental trials. Table~\ref{tab:fidelities} summarizes the results for the fidelities obtained. It is seen that the three-qubit state reconstructed from its two-qubit reduced marginals shows slightly larger fidelity than the state obtained from full three-qubit tomography, indicating the practical utility of the reduced-state approach. 

\begin{table}[h]
    \caption{Fidelity values between the theoretical state $\rho_{ABC}$ and the experimentally reconstructed states $\rho^{I}_{ABC}$ , $\rho^{II}_{ABC}$  for  five independent experimental trials.}
    \begin{tabular}{|c| c| c| c| c|}  
        \hline
        \multirow{2}{*}{\textbf{Trial No.}} & \multicolumn{2}{c|}{$F\left(\rho_{ABC},\rho^{I}_{ABC}\right)$} & \multicolumn{2}{c|}{$F\left(\rho_{ABC},\rho^{II}_{ABC}\right)$} \\ 
        \cline{2-5}
        & Unmitigated & Mitigated & Unmitigated & Mitigated \\ 
        \hline \hline
        1 & 0.8872 & 0.8917 & 0.9961 & 0.9970  \\ 
        \hline 
        2 & 0.9292 & 0.9330 &  0.9949 & 0.9965   \\ 
        \hline
        3 &  0.9321 & 0.9377 & 0.9892 & 0.9918  \\ 
        \hline
        4 &  0.8510 & 0.8547 &0.9678 & 0.9785  \\ 
        \hline
        5 & 0.9041 & 0.9128 & 0.9954 & 0.9957  \\ 
        \hline
    \end{tabular}
     \label{tab:fidelities}
\end{table}

\begin{figure}
\begin{subfigure}[t]{0.5\textwidth}
        \includegraphics[width=1.0\linewidth]{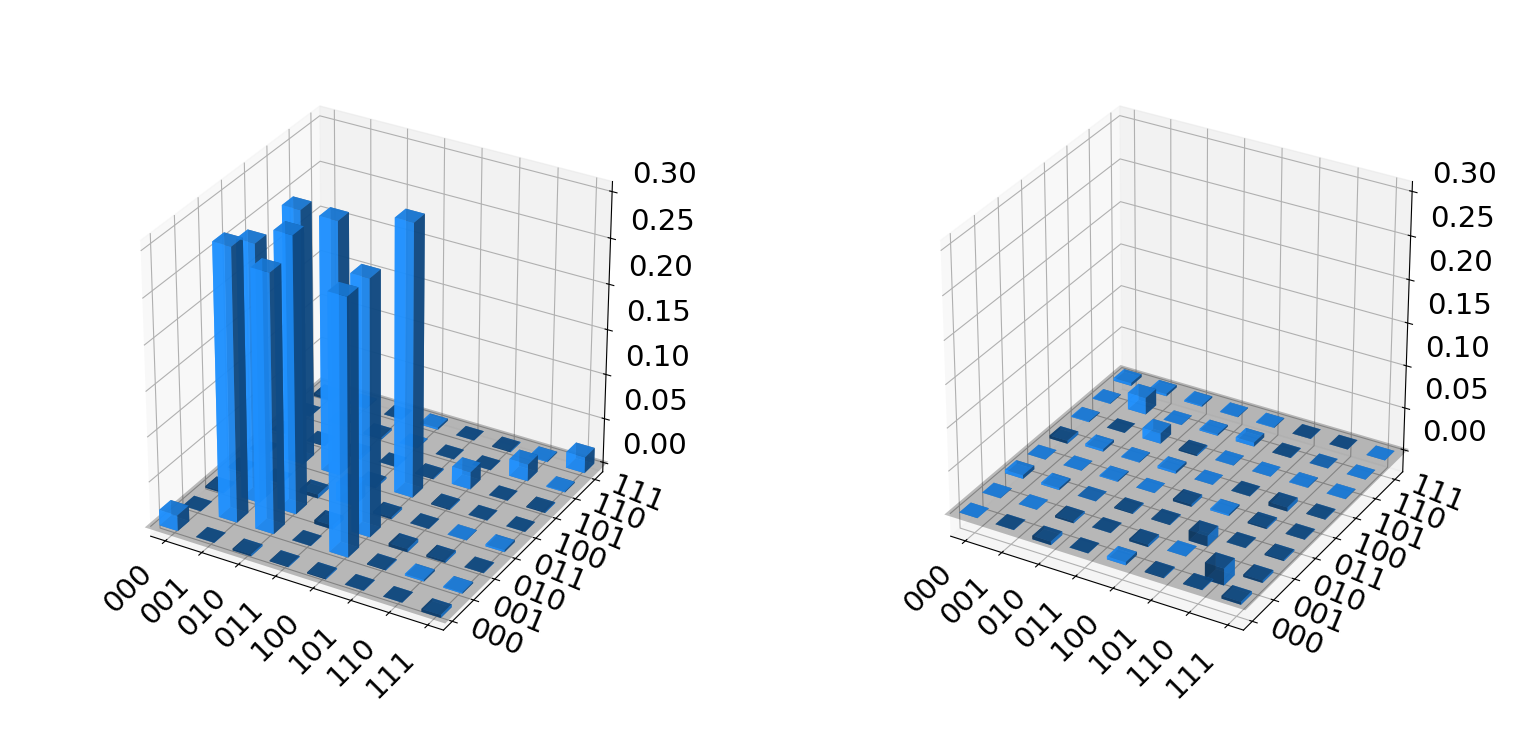}
    \caption{} \label{fig:e}
\end{subfigure}
\begin{subfigure}[t]{0.5\textwidth}
     \includegraphics[width=1.0\linewidth]{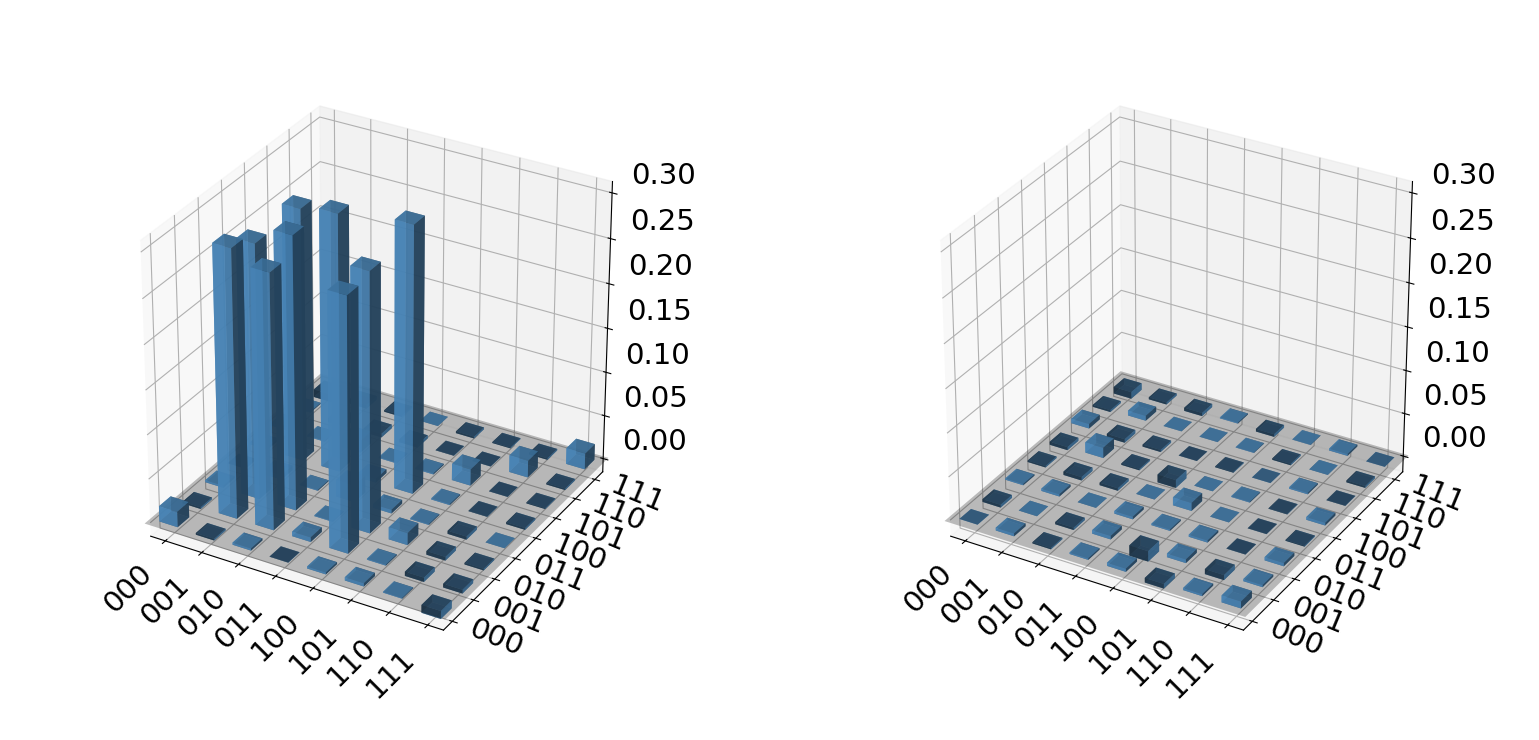}
     
    \caption{} \label{fig:f}
\end{subfigure}
 \caption{Real(left) and imaginary(right) parts of the elements of the MLE reconstructed density matrices:   (a) $\rho^I_{\rm MLE}$ generated from 17-setting full state tomography and (b) $\rho_{\rm MLE}^{II}$ obtained using the marginal two-qubit tomography.}
 \label{fig:mle}
\end{figure}

In order to assess the effectiveness of the reconstruction approaches, we have also implemented maximum likelihood estimation (MLE) ~\cite{james2001measurement, Kaznady_2009}. The MLE approach produces a bona fide density matrix by finding the positive semidefinite matrix that best fits the tomography scheme employed. We generated explicit three-qubit MLE density matrices $\rho^{I}_{\rm MLE}$, $\rho^{II}_{\rm MLE}$,  corresponding to both the  17-setting full tomography and the reduced two-qubit tomography schemes, respectively, by employing the noisy reference state 
\begin{equation}
\label{Wnoisy}
\rho^{noisy}_{\rm ABC}= 0.15\, \left(\frac{I\otimes I\otimes I}{8}\right)  + 0.85\, \vert{\rm W}_{ABC}\rangle \langle {\rm W}_{ABC}\vert
\end{equation}
as input. This ensured that the reconstructed MLE states $\rho^{I}_{\rm MLE}$, $\rho^{II}_{\rm MLE}$ are not rank-deficient, avoiding the issues of unjustified zero eigenvalues of outputs~\cite{Blume2010}. The resulting MLE density matrices for the full tomography and the marginal tomography schemes are given explicitly by  
\begin{widetext}
{\footnotesize \begin{equation*}
\rho^{I}_{\rm MLE} =
\begin{pmatrix}
0.01 & 0 & 0 & 0 & 0 & 0 & 0 & 0 \\
0 & 0.30 & 0.28 & 0 & 0.28  & 0 & -0.02i & 0 \\
0 & 0.28 & 0.30 & 0 & 0.28 & 0 & 0 & 0 \\
0 & 0 & 0 & 0.02 & 0 & 0 & 0 & 0 \\
0 & 0.28 & 0.28  & 0 & 0.30 & 0 & 0 & 0 \\
0 & 0 & 0.01i & 0 & 0 & 0.02 & 0 & 0 \\
0 & 0.02i & 0 & 0 & 0 & 0 & 0.01 & 0 \\
0 & 0 & 0 & 0 & 0 & 0 & 0 & 0.01
\end{pmatrix}, \ \ 
\rho^{II}_{\rm MLE} =
\begin{pmatrix}
0.01 & 0 & 0 & 0 & 0 & 0 & 0 & 0 \\
0 & 0.30 & 0.28 & 0 & 0.28-0.01i  & 0 & 0 & 0 \\
0 & 0.28 & 0.30 & 0 & 0.29 & 0.01 & 0 & 0 \\
0 & 0 & 0 & 0.01 & 0 & 0 & 0 & 0 \\
0 & 0.28+0.01i & 0.29  & 0 & 0.30 & 0 & 0 & 0 \\
0 & 0 & 0.01 & 0 & 0 & 0.01 & 0 & 0 \\
0 & 0.02i & 0 & 0 & 0 & 0 & 0.01 & 0 \\
0 & 0 & 0 & 0 & 0 & 0 & 0 & 0.01
\end{pmatrix}.
\end{equation*}}
\end{widetext}
 The corresponding bar plots of the MLE density matrices $\rho^{I}_{\rm MLE}$  and $\rho^{II}_{\rm MLE}$ are displayed in Fig.~\ref{fig:mle}. We find that the fidelities $F^I_{\rm MLE}~=~\left(\langle W_{ABC} \vert \rho^{I}_{MLE}\vert W_{ABC}\rangle \right)^{1/2}~=~0.9313$ for tomography with 17 measurement settings and $F^{II}_{\rm MLE}~=~\left(\langle W_{ABC} \vert \rho^{II}_{MLE}\vert W_{ABC}\rangle \right)^{1/2}~=~0.9359$ for marginal tomography scheme.  We performed a bootstrap error analysis~\cite{efron1994bootstrap} with 100 resampled MLE outputs for both tomography schemes.  For the  17-setting full tomography method, the mean fidelity is given by  $F^I_{\rm mean}=0.9313$ and standard deviation  $\sigma_I=0.0033$.  From the two-qubit marginal tomography, we obtained a mean fidelity $F^{II}_{\rm mean}$=0.9359 and standard deviation    $\sigma_{II}=0.0036.$ These fidelities are seen to be slightly lower than those obtained directly from experimental standard  tomography i.e., linear inversion followed by spectral correction approach (see Table~\ref{tab:fidelities}).  However, such reduction is expected because the  MLE procedure enforces physicality constraints on the density matrices and smooths out statistical fluctuations and noise present in finite data samples, leading to more reliable but sometimes less optimistic fidelity estimates~\cite{Blume2010,PRA2017}. 

\section{Summary}
\label{sec:Summay} 
We have  experimentally implemented a resource-efficient quantum state tomography scheme for generic three-qubit  states using IBM's \texttt{ibm\textunderscore{osaka}} superconducting quantum processor. A tomography protocol consisting of  17 measurement  settings  was developed and employed for reconstructing the full three-qubit density matrix of the three qubit W state. We have incorporated measurement error mitigation techniques and spectral correction to ensure physical validity of reconstructed density matrices. In addition to the full three-qubit state tomography, we have also employed a reduced-state approach based on a  two-qubit tomography scheme involving only seven measurement  settings  for reconstructing subsystem density matrices of the three-qubit state. Our experimental implementation involved {\em five} independent trials consisting of a total of 70,000 shots, ensuring robust data for both full- and reduced-state reconstructions. Using experimentally reconstructed two-qubit marginals $\rho^{\text{expt}}_{AB}$ and $\rho^{\text{expt}}_{BC}$, we successfully reconstructed the global three-qubit  state $\rho_{ABC}$.  While related reduced-state-based reconstruction of pure states has been demonstrated previously on differnt physical platforms~\cite{DDA15, Xin2017,Hu2024}, our work implements this protocol on an IBM quantum processor, serving as a proof-of-principle that global states can be inferred from partial tomographic information in a hardware-efficient and experimentally viable manner. 

We computed fidelities across five independent experimental trials. The reconstructions obtained from reduced two-qubit marginals consistently exhibited slightly higher fidelities compared to those from full three-qubit tomography. The lower fidelities observed in the 17-setting full-state reconstruction using  the standard quantum tomography methods, relative to that using 7 setting marginal tomography  scheme, can be attributed to hardware noise in current NISQ devices. The full tomography circuits entail a larger number of CNOT gates, increasing circuit depth and, consequently, SPAM and gate errors. In contrast, reduced-state circuits are shallower and less error-prone. 

To ensure physical validity and enable error analysis on fidelities, we employed MLE for both full and reduced tomography schemes using a noisy W state (see Eq.~(\ref{Wnoisy})) as the reference input, thereby generating robust output states and avoiding rank-deficiency pitfalls of the MLE approach~\cite{Blume2010}. The mean fidelity and standard deviation were calculated over 100 independent MLE reconstructions for each tomography scheme, providing a statistical characterization of the  MLE reconstruction accuracy. The mean fidelities obtained via MLE were found to be slightly lower than those obtained from experimental trials. This modest reduction could be attibuted to the enforcement of  physicality constraints in the MLE procedure and its tendency to smoothen out statistical fluctuations in finite data samples employed for simulation~\cite{Blume2010,PRA2017}.
\\

\section*{Acknowledgements}
  Sudha would like to acknowledge the funding under the  Q-Pragathi Project (No.~QP202406) of the Quantum Research Park, Karnataka Innovation and Technology Society (KITS), K-Tech, Government of Karnataka.  We acknowledge the use of IBM Quantum services for this work. The views expressed are those of the authors, and do not reflect the official policy or position of IBM or the IBM Quantum team.

%\section{References}
\bibliography{ref}
\end{document}